\documentclass[prl,aps,superscriptaddress,showpacs,reprint]{revtex4-1}

\usepackage{etoolbox}
\usepackage{gensymb}
\usepackage{hyperref}
\usepackage{amsmath}
\usepackage{amssymb}
\usepackage{dsfont}
\usepackage[T1]{fontenc}
\usepackage{ae}
\usepackage{color}
\usepackage{graphicx} 
\usepackage{units}
\usepackage{placeins}

\begin{document}

\title{Single-shot Non-destructive Detection of Rydberg Atom Ensembles \\ by Transmission Measurement of a Microwave Cavity}

\author{S. Garcia} \email{Present address: Institute of Physics, Coll{\`e}ge de France, Paris, France, sebastien.garcia@college-de-france.fr} \affiliation{Department of Physics, ETH Z\"urich, CH-8093 Z\"urich, Switzerland}

\author{M. Stammeier}  \affiliation{Department of Physics, ETH Z\"urich, CH-8093 Z\"urich, Switzerland}

\author{J. Deiglmayr} \email{Present address: University of Leipzig, Felix-Bloch Institut, Leipzig, Germany} \affiliation{Laboratorium f\"ur Physikalische Chemie, ETH Z\"urich, CH-8093, Z\"urich, Switzerland} 

\author{F. Merkt} \affiliation{Laboratorium f\"ur Physikalische Chemie, ETH Z\"urich, CH-8093, Z\"urich, Switzerland}

\author{A. Wallraff} \affiliation{Department of Physics, ETH Z\"urich, CH-8093 Z\"urich, Switzerland}

\begin{abstract}
We present an experimental realization of single-shot non-destructive detection of ensembles of helium Rydberg atoms. We use the dispersive frequency shift of a superconducting microwave cavity interacting with the ensemble. By probing the transmission of the cavity, we determine the number of Rydberg atoms or the populations of Rydberg quantum states when the ensemble is prepared in a superposition. At the optimal microwave probe power, determined by the critical photon number, we reach single-shot detection of the atom number with $13\%$ relative precision for ensembles of about $500$ Rydberg atoms with a measurement backaction characterized by approximately $2\%$-population transfer. 

\end{abstract} 	
\maketitle 

Rydberg atoms have extreme properties~\cite{Gallagher1994,Saffman2010} which are used in many experiments investigating quantum computation~\cite{Jaksch2000,Isenhower2010,Saffman2010}, quantum simulation~\cite{Schauss2015,Labuhn2016,Bernien2017}, quantum optics~\cite{Haroche2013,Peyronel2012,Maxwell2013a,Baur2014,Liang2018} or chemical reaction dynamics~\cite{Bendkowsky2009,Schlagmuller2016,Allmendinger2016a}. The standard technique for detection of Rydberg atoms relies on field ionization followed by counting of ions or electrons using an avalanche detector~\cite{Gallagher1994}. More recently, experiments with arrays of trapped single atoms~\cite{Schauss2015,Labuhn2016,Bernien2017} detect Rydberg excitations by ground-state atom loss with fluorescence imaging. Both methods are efficient but destructive. Non-destructive detection offers tools for quantum-state engineering, in particular with characterization of the initial state, measurement-induced multipartite entanglement~\cite{SchleierSmith2010,Haas2014}, quantum Zeno effect~\cite{Bernu2008,Barontini2015} and quantum feedback~\cite{Sayrin2011,Vijay2012}. Rydberg atoms can be detected non-destructively with optical photons using electromagnetically-induced transparency~\cite{Gunter2012,Gunter2013,Gorniaczyk2014}. However, this technique relies on the change of the optical response of a dense gas of ground-state atoms within the Rydberg-blockade volume surrounding a Rydberg atom. Thus, it remains impractical for spatially extended ensembles of Rydberg atoms such as atomic beams or large single-atom arrays.

Cavity-quantum-electrodynamics (QED) techniques allow efficient single-shot non-destructive measurements for many different systems including atoms~\cite{Volz2011a} and superconducting qubits~\cite{Jeffrey2014,Walter2017}. In this context, Rydberg atoms have been used to probe non-destructively microwave photons confined in a cavity~\cite{Nogues1999,Haroche2013}. This detection relies on the photon-number-dependent dispersive shift of the Rydberg levels induced by the non-resonant atom-photon interaction, which also leads to a dispersive shift of the cavity frequency induced by the atoms. The latter effect was used in Ref.~\cite{Maioli2005} to determine the Rydberg-atom number or state, but the presented method required extremely long-lived cavity photons, measured with additional destructively-detected Rydberg atoms.
In contrast, cavity-transmission measurements, used in optical cavity QED and circuit QED, offer direct detection. 

In this letter, we present experiments in which we reach the single-shot detection regime, where the uncertainty~\cite{BIPM2012} of a single measurement becomes much smaller than the signal, for Rydberg-atom ensembles by measuring microwaves transmitted through a cavity. To this end, we combine a cavity offering a high sensitivity to the dispersive shift with an optimized signal-to-noise ratio (SNR) of the detection.
By using a superconducting cavity with high internal quality factor of $1.7 \cdot 10^6$, we obtain an increase in sensitivity to the Rydberg atom number by a factor of $20$ over our previous results~\cite{Stammeier2017}. 
We demonstrate that transmission measurements enable the detection of the populations of Rydberg quantum states coupled to the cavity when the ensemble is prepared in a superposition state. By optimizing the SNR with a microwave probe power corresponding to the critical photon number, we achieve single-shot detection with a relative precision of $13\%$ (statistical standard deviation) for Rydberg ensembles of about $500$ atoms. This presents an opportunity for applications in experiments with ensembles of Rydberg atoms, in particular for detection of Rydberg atoms in cold-collision experiments or for quantum-state engineering in atomic-array experiments.

\begin{figure}[t] \centering \includegraphics[width=80mm]{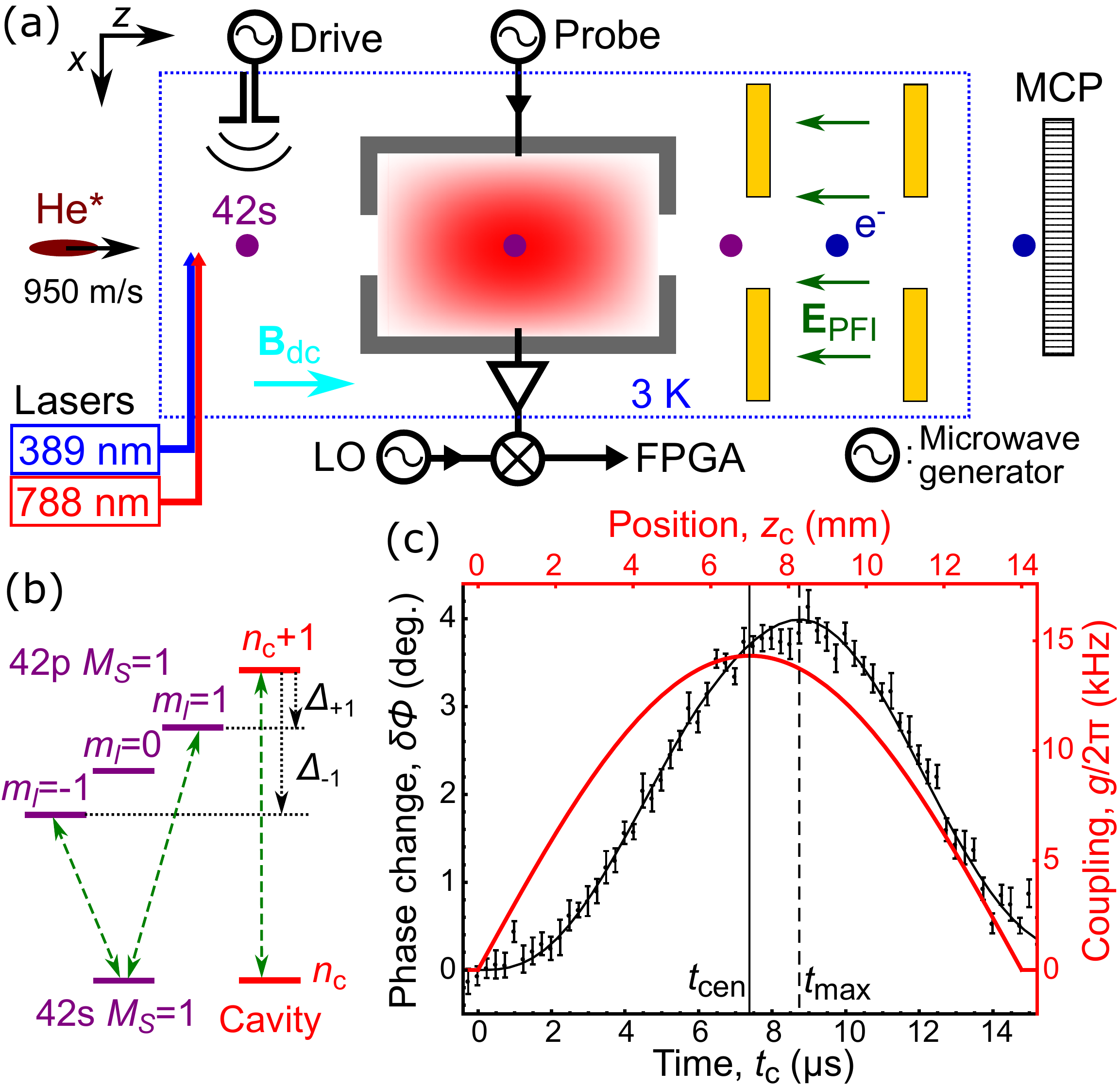} 	
\caption{(a) Experimental setup combining the generation of an ensemble of He Rydberg atoms, its non-destructive detection by transmission measurement of a microwave cavity and its destructive detection by field ionization followed by a measurement of the number of extracted electrons on a MCP. (b) Energy-level diagram of the Rydberg atom and the cavity, with dashed arrows marking the coupled transitions and dotted arrows indicating the atom-cavity detunings $\Delta_{\pm1}$. (c) Single-atom-cavity coupling strength $g$ (thick curve, right axis) and measured phase change for a microwave probe frequency on resonance with the cavity (data points and fitted model, left axis) versus the elapsed time since the atoms entered the cavity $t_{\mathrm{c}}$ and equivalently versus their position $z_{\mathrm{c}}=t_{\mathrm{c}}\cdot v$. The data are averaged over $5.5\cdot 10^4$ repetitions and error bars indicate $\pm$ the standard uncertainty. Vertical full and dashed lines indicate the time $t_{\mathrm{cen}}$ at which the atoms reach the cavity center and $t_{\mathrm{max}}=t_{\mathrm{cen}}+2/\kappa$, respectively.}
\label{fig:setup} 
\end{figure}	

Our experimental setup, schematically shown in Fig.~\ref{fig:setup}(a), combines a non-destructive detection by measurement of the microwave-cavity transmission with a subsequent destructive detection by ionization. The experiment uses an atomic beam pulsed with a repetition rate of $25\,$Hz. We start each measurement cycle by creating a cloud of metastable ($1\mathrm{s}^{1} \, 2\mathrm{s}^{1} \,^{3}\mathrm{S}_{1}$) helium atoms, which propagates at a mean velocity $v=950(20)\, \mathrm{m}\cdot\mathrm{s}^{-1}$ through a cryogenic environment at $3\,$K~\cite{Thiele2014,Stammeier2017}. We prepare a small cloud of atoms in the triplet $42\mathrm{s},\, M_S=1$ Rydberg state, by two-photon excitation via the $3\, ^{3}\mathrm{P}_{0}$ state in a magnetic field of $\sim 5\,$G. At this field strength, the atomic structure is in the Paschen-Back regime, with approximate quantum numbers $M_S$ and $m_l$. The atoms then interact with the microwave photons confined in a superconducting rectangular 3D cavity~\cite{Stammeier2017a}. We measure the change of the complex transmission (amplitude and phase) of a microwave probe tone through the cavity, which results from the dispersive shift $\chi$ of the cavity frequency induced by the Rydberg atoms. The transmitted microwave tone is amplified and down-converted to a $25$-MHz intermediate frequency and recorded using an FPGA-based device (field-programmable gate array)~\cite{Stammeier2017}. After the cloud has interacted with the cavity, we perform a state-selective destructive detection  by pulsed-field ionization followed by detection of the extracted electrons on a micro-channel plate detector (MCP), see Sup.~Mat.~A.

The cavity-atom interaction is characterized by (i) the single atom-photon coupling strength $g$, which is proportional to the microwave electric field and reaches $g_{\mathrm{max}}/2\pi \simeq 14.3 \,$kHz at the cavity center (see Fig.~\ref{fig:setup}(c)), (ii) the number $N$ of Rydberg atoms, and (iii) the detunings $\Delta_{\pm 1} = \omega_{\pm 1} -\omega_{\mathrm{c}}$ between the cavity frequency ($\omega_{\mathrm{c}}/2\pi \simeq 20.5583\,$GHz) and the frequencies ($\omega_{\pm 1}/2\pi$) of the transitions to the $42\mathrm{p},\, M_S=1,\, m_l=\pm 1$ levels (see Fig.~\ref{fig:setup}(b)). Because the cavity microwave electric field is orthogonal to the magnetic quantization field, the transition to the $\mathrm{p},\, m_l=0$ state is not coupled to the cavity mode. The detunings are determined by the strength of the applied magnetic field~\cite{Stammeier2017a} and are measured in-situ using microwave spectroscopy ($\Delta_{+1}/2\pi \simeq -8\,$ MHz and $\Delta_{-1}/2\pi \simeq -26\,$ MHz for the data in Fig.~\ref{fig:setup}(c)).
The detunings are much larger than the collective coupling ($\mathrm{max}(g\sqrt{N})/(2\pi) \sim 0.4\,$ MHz), and therefore the interaction leads to a dispersive shift $\chi$ of the cavity frequency. By calculating the eigenvalues of the Tavis-Cummings Hamiltonian~\cite{Tavis1968}, extended to a 3-level-atom ensemble and a single-mode cavity, up to second order in $g\sqrt{N}/\Delta_{\pm1}$ (with $g\sqrt{N}\ll(\Delta_{+1}-\Delta_{-1})$), the cavity dispersive shift
\begin{equation}
\chi=g^2N\left[\frac{1}{\Delta_{+1}}\left(P_{\mathrm{p},+1}-P_{\mathrm{s}}\right)+\frac{1}{\Delta_{-1}}\left(P_{\mathrm{p},-1}-P_{\mathrm{s}}\right)\right] 
\label{eq:chi}
\end{equation}
is the sum of two terms linked to two atomic transitions (with identical coupling $g$), where $P_{\mathrm{s}}$ and $P_{\mathrm{p},\pm1}$ are the fraction of atoms in the $42$s and the $42$p $m_{\mathrm{l}}=\pm 1$ states, respectively. 

Measuring the phase change $\delta\phi$ on resonance provides the best sensitivity $\mathrm{d}\delta\phi/\mathrm{d}\chi$ to detect a dispersive shift.
Indeed, the phase change induced by a constant dispersive shift $\chi$ on a resonant microwave probe tone is given by $\delta\phi=-\arctan(2 \chi/\kappa)$ where $\kappa/2\pi=236(1)\,$kHz is the cavity linewidth. Thus, $\delta\phi$ is approximately linear in $\chi$ for small shifts. A resonant microwave probe frequency also maximizes the transmitted power and consequently the precision of the phase detection.

\begin{figure}[t] \centering \includegraphics[width=85mm]{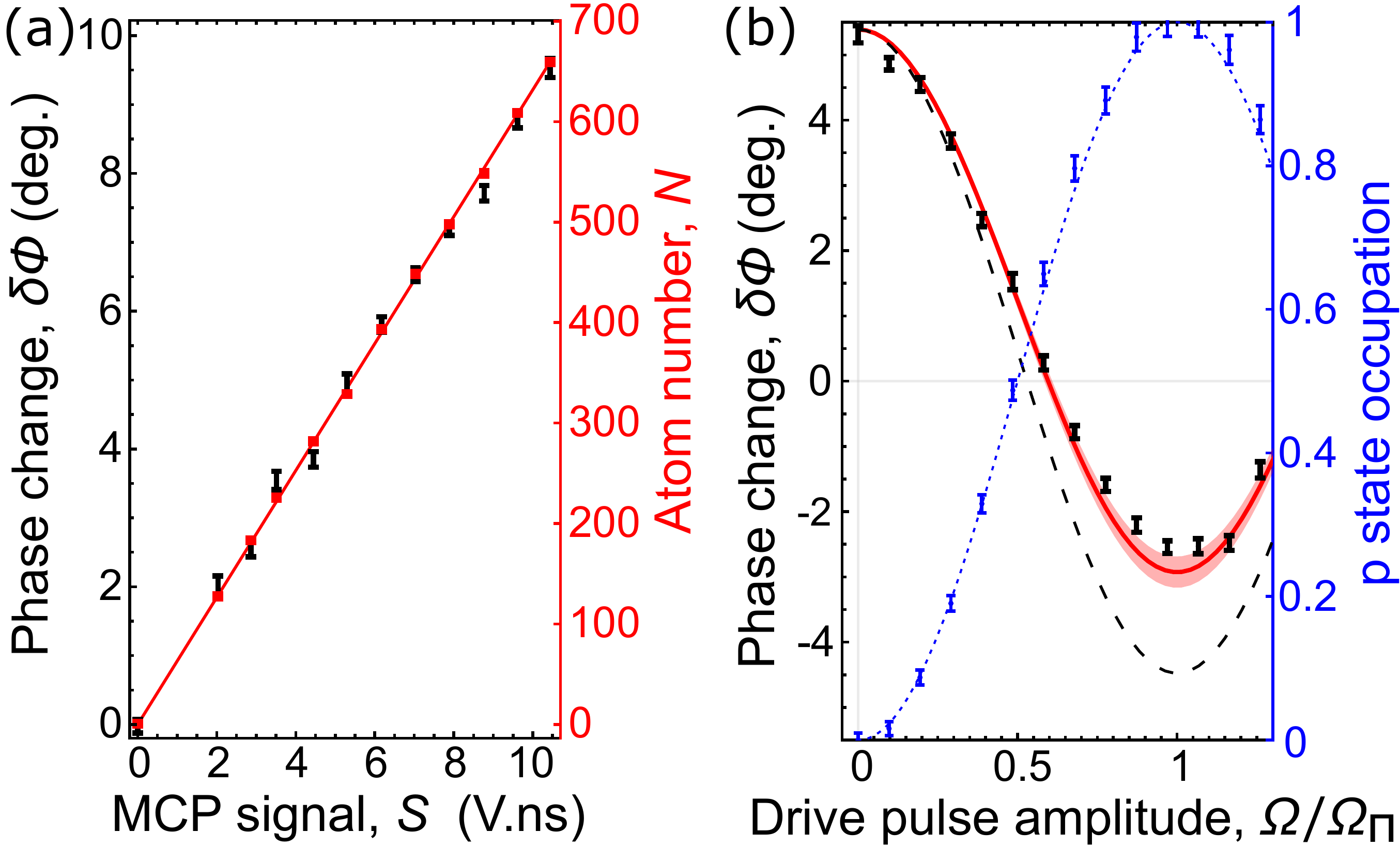} 	
\caption{(a) Phase change $\delta\phi$ at $t_{\mathrm{max}}$ (black, averaged over $1\,\mu$s and $3.8\cdot 10^4$ repetitions) and extracted atom number (red) versus MCP signal $S$ when varying the Rydberg-atom number $N$ produced by laser excitation. The linear fit (red line) emphasizes the proportionality between $S$ and $\delta\phi$. (b) Phase change at $t_{\mathrm{max}}$ (black, averaged over $1\,\mu$s and $4.3\cdot 10^4$ repetitions) and p-state occupation (blue) as a function of the normalized Rabi frequency of the microwave excitation from the s state to the p$,\, m_l=+1$ state. The dotted curve is the expected Rabi oscillation of the p state occupation. The dashed curve is the phase change calculated for an atom cloud in a pure  p$,\, m_l=+1$ state in the cavity. The full  curve shows the calculation including the independently measured intra-cavity populations of p-state sublevels with uncertainty depicted by the shaded region. Error bars indicate $\pm$ the standard uncertainty.}
\label{fig:AtomPol} 
\end{figure}

We observe the average phase change of the cavity transmission when a cloud of $42$s state atoms ($P_{\mathrm{s}}=1$, $P_{\mathrm{p},\pm1}=0$) propagates through the cavity, see Fig.~\ref{fig:setup}(c). The change of the coupling strength along the cloud's path induces a variation of the dispersive shift, because $\chi \propto g^2$. The cavity field reacts to this variation with a response function characterized by the time constant $\tau_{\mathrm{c}}=2/\kappa\simeq1.35\,\mu$s. Thus, we observe a shift between the time $t_{\mathrm{cen}}$, at which the atoms reach the cavity center, and the time of the maximum phase change at approximately $t_{\mathrm{max}}=t_{\mathrm{cen}}+\tau_{\mathrm{c}}$. 
With the calculated coupling and the measured detunings, we fit the measured cavity transmission using input-output theory~\cite{Gardiner1985} (details in Sup.~Mat.~B). 
We extract the average atom number ($N=261(1)$) in the center of the cavity from the fit showing excellent agreement with the data (see Fig.~\ref{fig:setup}(c)),
\FloatBarrier

To characterize the sensitivity $\mathrm{d}\delta\phi/\mathrm{d}N$ of the cavity detection to the number of Rydberg atoms, we generate $42$s atom clouds with different average numbers of Rydberg atoms by changing the duration of the $389$-nm-laser pulse from $0$ to $1.1\,\mu$s. In Fig.~\ref{fig:AtomPol}(a), we present the correlation between the integrated field-ionization signal $S$ of the MCP and both the phase change $\delta\phi$ on resonance at $t_{\mathrm{max}}$ and the extracted number of atoms $N$. Because the dispersive shift $\chi$ is smaller than the cavity linewidth $\kappa$, the phase change is proportional to the dispersive shift and thus to the atom number. Our cavity Rydberg atom detector has a phase sensitivity $\mathrm{d}\delta\phi/\mathrm{d}N=1.44(1)\cdot 10^{-2}\,$degree$/$atom. This $20$-fold improvement over previously published results~\cite{Stammeier2017} originates from the smaller linewidth of the superconducting cavity. The cavity linewidth is limited by the external coupling to maximize the SNR and to avoid a cavity-induced averaging timescale $\tau_{\mathrm{c}}$ longer than the interaction time, which would reduce the sensitivity. 

The MCP signal $S$ is proportional to the number of atoms. We independently determined the MCP sensitivity $\mathrm{d}S/\mathrm{d}N=2.07(4)\cdot 10^{-2}\,\mathrm{V}\cdot\mathrm{ns}/$atom from the measured single-electron response and the known MCP detection efficiency ($\eta \simeq 55\%$). When accounting for the radiative decay of the Rydberg atoms between the cavity center and the ionization region, the fitted linear relation between the MCP signal and the number of atoms detected with the cavity measurement (see Fig.~\ref{fig:AtomPol}(a)) also provides a calibration of the MCP sensitivity of $2.12(1)\cdot 10^{-2}\,\mathrm{V}\cdot\mathrm{ns}/$atom. From the difference between the two independent calibrations, we deduce the trueness of the atom-number detection by the cavity, characterized by a relative systematic error of $-2.3(1.8)\%$, consistent with a $-2.4(0.8)\%$ estimate from the approximations of the model (see Sup.~Mat.~C, which includes Refs.~\cite{Zimmerman1979,Drake1999,Pozar2011,Zhelyazkova2016}). We conclude that the cavity measurement provides an accurate non-destructive detection of the number of Rydberg atoms.

Cavity transmission measurements also provide a non-destructive detection of the populations of quantum states for atoms in optical cavities~\cite{Volz2011a} or superconducting qubits in microwave cavities~\cite{Wallraff2005} for example. To measure the state populations of a Rydberg ensemble in the microwave cavity, we prepare the ensemble in different superpositions of the s and p$,\,m_l=+1$ states before the atoms enter the cavity, by applying a microwave drive pulse with amplitude characterized by the Rabi frequency $\Omega$. The p-state occupation, detected with state-selective ionization after the cavity (Fig.~\ref{fig:AtomPol}(b)), follows the expected Rabi oscillation, which indicates that the cavity-measurement backaction is limited to the projection onto the eigenstates of the p-state occupation.
The observed phase change reflects this sinusoidal variation of the populations. It reaches minimal values when the ensemble is prepared in the p state with a $\pi$-pulse ($\Omega=\Omega_{\pi}=\pi/\delta t$ with $\delta t=0.4\,\mu$s the pulse duration), because this state induces a dispersive shift opposite to that of the s state. 
However, the amplitude of the phase change is reduced as compared to the initial s-state ensemble, because only one transition contributes to the dispersive shift for p$,\,m_l = +1$ (see Eq.~(\ref{eq:chi}), dashed line). Additionally, a depolarization of the ensemble, occurring between the location of the state preparation and the cavity, transfers a part of the p-state population from $m_l=+1$ to $m_l=0$ and $m_l=-1$. We attribute this depolarization to stray electric fields at the cavity entrance hole, which could be avoided with an improved cavity design. Using in-situ microwave spectroscopy, we determined the resulting populations of the p-state sublevels to be $P_{\mathrm{p},+1}=0.61(3)$ and $P_{\mathrm{p},-1}=0.20(3)$ (see Sup.~Mat.~D). The corresponding calculated response (red) agrees with the observed phase change. Thus, the microwave-cavity-transmission measurement enables the non-destructive detection of the states populations of a Rydberg ensemble.

\FloatBarrier

\begin{figure}[t] \centering \includegraphics[width=85mm]{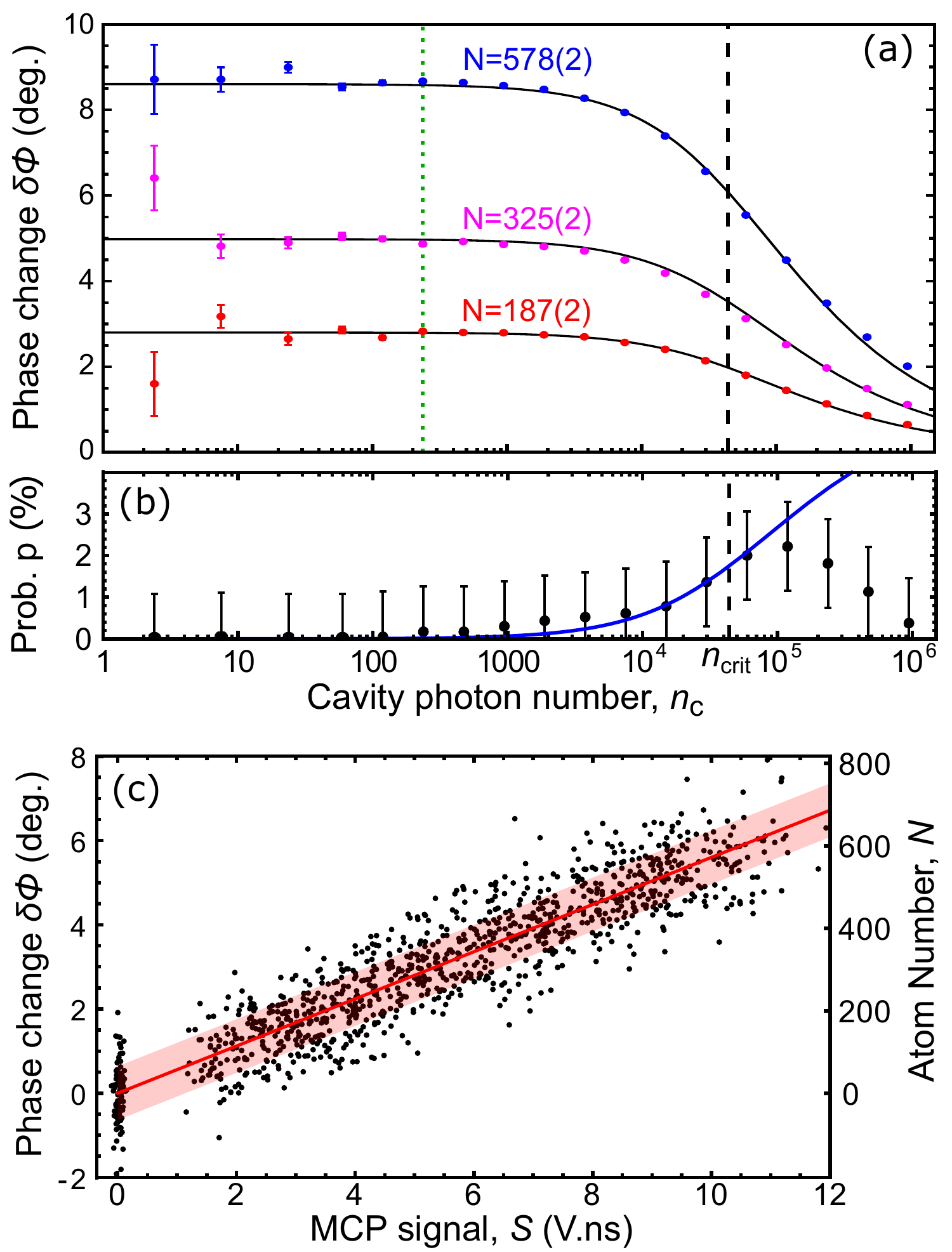} 	
\caption{(a) Phase change $\delta\phi$ (at $t_{\mathrm{max}}$ averaged over $\tau_{\mathrm{i}}=6.2\,\mu$s and $3\cdot 10^4$ repetitions) as a function of the number of cavity photons $n_{\mathrm{c}}$ for the indicated average numbers of Rydberg atoms. The global fit (solid lines) determines the critical photon number (dashed vertical line). The dotted vertical line indicates the photon number used in the measurements presented in Fig.~\ref{fig:AtomPol}(b). (b) Probability of excitation to p state as a function of $n_{\mathrm{c}}$, measured using the MCP. Error bars are dominated by systematic uncertainties. The solid line is a fit of the data to the model presented in the text. (c) Single-shot measurements (dots) of the phase change $\delta \phi$, and deduced Rydberg-atom number $N$, vs. the MCP signal $S$. The average number of Rydberg atoms is controlled during the preparation by the pulse duration of the laser excitation ($0$ to $1\,\mu$s in $0.1\,\mu$s steps).  The shaded region indicates the standard deviation $\sigma_N\simeq 65$ of the atom number extracted from the transmission measurement.}
\label{fig:SingleShot} 
\end{figure}

The main interest of non-destructive detection lies in its single-shot-measurement abilities. In order to reach the single-shot regime for the detection of the Rydberg atom number, additionally to the high sensitivity presented above, we optimized the SNR. The power SNR of the microwave detection $R_{\mathrm{S/N}}=n_{\mathrm{c}}\kappa_{\mathrm{out}} \tau_{\mathrm{i}}/n_{\mathrm{noise}}$ depends on the number of photons in the cavity $n_{\mathrm{c}}$ (proportional to the microwave probe power, calculated using input-output theory), the coupling of the output port $\kappa_{\mathrm{out}}/2\pi\simeq 150\,$kHz, the integration time $\tau_{\mathrm{i}}$ and the effective noise photon number $n_{\mathrm{noise}}\simeq 23$ of the detection, which is dominated by the noise of the first amplifier. Thus, increasing $n_{\mathrm{c}}$ reduces the phase-detection precision $\sigma_\phi=1/\sqrt{R_{\mathrm{S/N}}}$. However, simultaneously the dispersive shift decreases as $\chi\left(n_{\mathrm{c}}\right)=\chi_0/\sqrt{1+n_{\mathrm{c}}/n_{\mathrm{crit}}}$ where the critical photon number $n_{\mathrm{crit}}=\Delta^2/(4g^2)$ characterizes the threshold of the validity of the dispersive approximation~\cite{Blais2004} and $\chi_0$ is the dispersive shift for $n_{\mathrm{c}}\ll n_{\mathrm{crit}}$ given by Eq.~(\ref{eq:chi})~\cite{Stammeier2017}. In Fig.~\ref{fig:SingleShot}(a), we present measurements of the dependence of the phase change on the photon number for three different average atom numbers. The global fit of the three data sets confirms that the measured power dependence is not influenced by the atom number and yields a critical photon number of $n_{\mathrm{crit}}\simeq 4.4\cdot 10^4$, consistent with the value calculated from the detuning $\Delta_{+1}$ and coupling $g$ averaged over the integration time. Due to the power dependences of the SNR and the dispersive shift, for $n_{\mathrm{c}}\gg n_{\mathrm{crit}}$, both the phase detection precision ($\sigma_\phi=1/\sqrt{R_{\mathrm{S/N}}}$) and the phase change ($\delta\phi\propto\chi$) scale as $1/\sqrt{n_{\mathrm{c}}}$. Thus, the atom-number precision only significantly improves with increasing power up to $n_{\mathrm{crit}}$.

Large photon numbers also induce a residual excitation of the atomic ensemble after the interaction with the cavity, which we measure with the MCP (see Fig.~\ref{fig:SingleShot}(b)). Such backaction is commonly observed in dispersive readout and arises from relaxation and dephasing~\cite{Boissonneault2009} during the interaction with the microwave probe field which mixes the atomic states. The atomic excitation of the system eigenstate in the cavity is $P_{\mathrm{e}}=\sin^2(\arctan[\sqrt{(n_{\mathrm{c}}+1)/n_{\mathrm{crit}}}])$~\cite{Walter2017,Narducci1973,Garraway2011}. As shown by the fitted curve, the residual excitation is proportional to $P_{\mathrm{e}}$ up to $n_{\mathrm{crit}}$. At larger microwave probe powers, the dispersive approximation breaks down and ac Stark shifts induced by couplings to other Rydberg states become significant; these effects hinder predictions of the residual excitation beyond $n_{\mathrm{crit}}$. The residual excitation remains below $\sim 2\%$ up to $n_{\mathrm{crit}}$, so that the detection can still be considered non-destructive. 

To achieve high precision with limited backaction, we perform single-shot measurements of the Rydberg-atom number with a cavity photon number ($n_{\mathrm{c}}\simeq 5.9\cdot 10^4$) close to the critical photon number. We integrate the signal over $\tau_{\mathrm{i}} = 6.2\,\mu$s, which corresponds to the time span over which $g\geq 0.8\,g_{\mathrm{max}}$. We observe the expected linear relationship between the single-shot results of the destructive detection by ionization and the non-destructive detection by measuring the phase change of the cavity transmission for different average Rydberg atom numbers, see Fig.~\ref{fig:SingleShot}(c). The scatter of the data points around the average response (full line) is caused by the precisions of the two detection methods. The precision of the phase detection limits the precision of the atom-number determination to a statistical standard deviation $\sigma_{N}\simeq 65$. Our non-destructive detector thus reaches a single-shot relative precision $\sigma_{N,\mathrm{rel}}=\sigma_{N}/N\simeq 13\%$ for ensembles of about $500$ atoms, which is on the same order of precision as the ionization detection (see Sup.~Mat.~E). 

The cavity-transmission detection has potential for significant improvement. For example, the noise of the microwave detection could be reduced to an effective noise photon number $n_{\mathrm{noise}}\sim 1$ by using a quantum-limited amplifier~\cite{Yurke1996, Castellanos2008}, leading to a relative precision of $4.8\%$ for $500$ atoms. The same improvement could be reached by increasing the integration time up to $\tau_{\mathrm{i}}\sim 0.1\,$ms, which could be obtained by decelerating the Rydberg atom cloud~\cite{Hogan2012a}.

In conclusion, we demonstrated non-destructive detection of atom number and atomic-state populations of Rydberg atom ensembles by microwave cavity transmission. Single-shot detection is achieved with small backaction of the measurement. The presented detection scheme opens up new perspectives for experiments with Rydberg ensembles. In particular, we anticipate that single-atom precision could be obtained with integration times on the order of $10\,$ms in experiments with trapped circular Rydberg atoms~\cite{Nguyen2018}.

\ 

We thank Tobias Thiele for his contributions to the initial phase of the experiment. This work was supported by the European Union H$2020$ FET Proactive project RySQ (grant N. $640378$) and by the National Centre of Competence in Research "Quantum Science and Technology" (NCCR QSIT), a research instrument of the SNSF.

\FloatBarrier

%

\newpage

\ 

\newpage

\appendix

{\LARGE Supplemental material}

\section{A. State selective detection by ionization}
\label{app:StateDet}

\begin{figure}[t] \includegraphics[width=80mm]{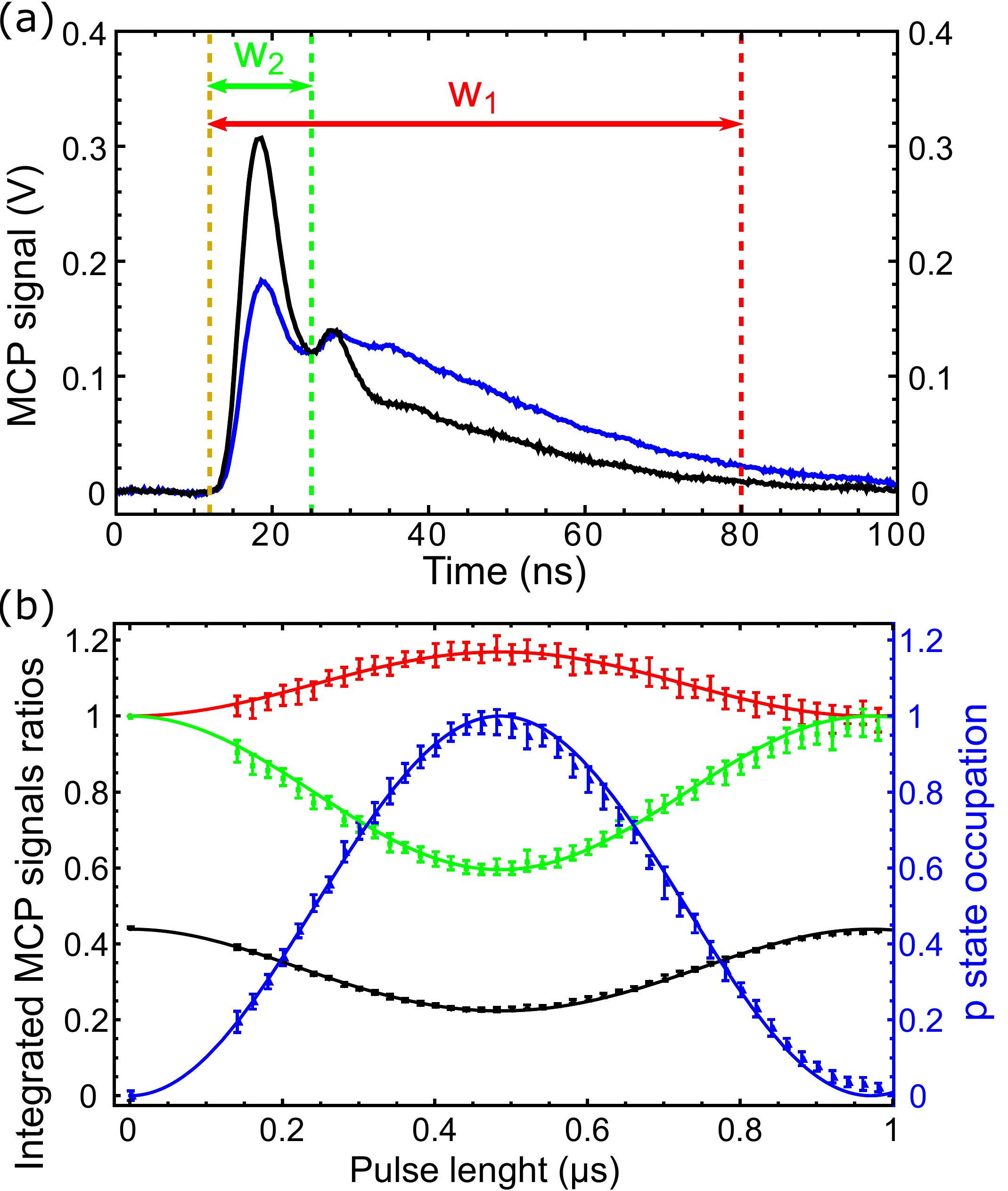} 	
\caption{(a) MCP signal measured after preparation of atoms in the $42$s state (black) or the $42$p state (blue). Vertical dashed lines and arrows indicate the two integration windows. (b) Rabi oscillations between s and p states measured by varying the microwave pulse duration. The integrated signals $S_1$ and $S_2$ (normalized to the $42$s-state values) over the windows W$_1$ and W$_2$  are presented in red and green, respectively. The black data points show the ratio $S_{\mathrm{r}} = S_2/S_1$. The corresponding full lines are a fit of the model (see text) from which we deduce the p state occupation (blue data points) which follows the expected $\sin^2$ dependence (blue) for Rabi oscillations.}
\label{fig:StateDet} 
\end{figure}

We measure the fraction of p-state atoms by destructive state-selective pulsed-field ionization. This measurement allows us to determine the atomic transition frequencies by microwave spectroscopy, to characterize the preparation of superposition states, and to quantify the backaction of the cavity detection on the population. To do so, we use the difference in MCP signal between the s and p states. In our pulsed-field-ionization approach, the s state ionizes faster and thus leads to larger signals at earlier times. We use two integration windows W$_1$ and W$_2$ capturing the full integrated signal $S_1$ and its first part $S_2$, see Fig.~\ref{fig:StateDet}(a). For a detected cloud containing $N_{\mathrm{s}}$ $42$s-state atoms and $N_{\mathrm{p}}$ $42$p-state atoms, the full signal is $S_1 = S_{1,\mathrm{a}} (N_{\mathrm{s}} + \alpha_{\mathrm{p}} N_{\mathrm{p}})$ where $S_{1,\mathrm{a}}$ is the integrated single-atom signal for an s-state atom and $\alpha_{\mathrm{p}}$ a relative coefficient for the signal of a p-state atom. The latter is close to one and reflects the fraction of the slowly decaying signal outside of the integration window. The signal of the first window is described by $S_2 = S_{1,\mathrm{a}} (\beta_{\mathrm{s}} N_{\mathrm{s}} + \alpha_{\mathrm{p}} \beta_{\mathrm{p}} N_{\mathrm{p}})$, where the coefficients $\beta_{\mathrm{s}}$ and $\beta_{\mathrm{p}}$ represent the relative contributions of $S_2$ to the full signal $S_1$ for an s-state and a p-state cloud, respectively.  

In order to use the difference in MCP signal to measure the fraction of p-states atoms in the cloud, we calibrate the coefficients $\alpha_{\mathrm{p}}$, $\beta_{\mathrm{s}}$ and $\beta_{\mathrm{p}}$. To this end, we measure $S_1$ and $S_2$ while changing the population from s-state to p-state with a microwave pulse applied after the laser excitation, see Fig.~\ref{fig:StateDet}(b) showing the signals normalized to the s-state signal. We clearly observe a sinusoidal variation of $S_2$ (green) resulting from the Rabi oscillation between s and p state. $S_1$ also shows a sinusoidal variation, but to larger values. This effect is mainly caused by the difference of lifetimes $\tau_{\mathrm{s}} \simeq 57.2\,\mu$s and $\tau_{\mathrm{p}} \simeq 102.6\,\mu$s of the s and p states and the difference in time $\Delta t_{\mathrm{m,d}} \simeq 35.5\,\mu$s between the microwave-induced population change and the ionization. 
Both $S_1$ and $S_2$ are proportional to the number of atoms $N$. Consequently, the ratio $S_{\mathrm{r}} = S_2/S_1$ is independent of $N$ and thus has the advantage of being insensitive to fluctuations of the number of prepared Rydberg atoms, as can be seen on the reduced uncertainty of the corresponding black data set in Fig.~\ref{fig:StateDet}(b). By fitting all three data sets jointly (full lines) taking into account radiative decay, we obtain the coefficients $\alpha_{\mathrm{p}} = 0.888(13)$, $\beta_{\mathrm{s}} = 0.439(3)$ and $\beta_{\mathrm{p}} = 0.222(3)$. 

With the calibrated coefficients, we determine the fraction $P_{\mathrm{p}}$ of atoms transferred from the s to the p state by measuring the ratio $S_{\mathrm{r}}$:
\begin{equation}
P_{\mathrm{p}} = \frac{1}{1+ \alpha_{\mathrm{p}} \frac{\beta_{\mathrm{p}} - S_{\mathrm{r}}}{S_{\mathrm{r}}-\beta_{\mathrm{s}}} e^{\Delta t_{\mathrm{m,d}} (\frac{1}{\tau_{\mathrm{s}}} - \frac{1}{\tau_{\mathrm{p}}})}} \ .
\end{equation}
The p-state occupation calculated from the ratio $S_{\mathrm{r}}$ is shown in Fig.~\ref{fig:StateDet}(b) and corresponds well to the expected sinusoidal Rabi oscillation (blue data set). The results presented in Fig.~3(b) of the main text and the characterization of the polarization of the ensemble inside the cavity use this method to determine the p-state population.

\FloatBarrier
\section{B. Temporal response of the cavity}
\label{app:TimeRes}

\begin{figure}[t] \includegraphics[width=80mm]{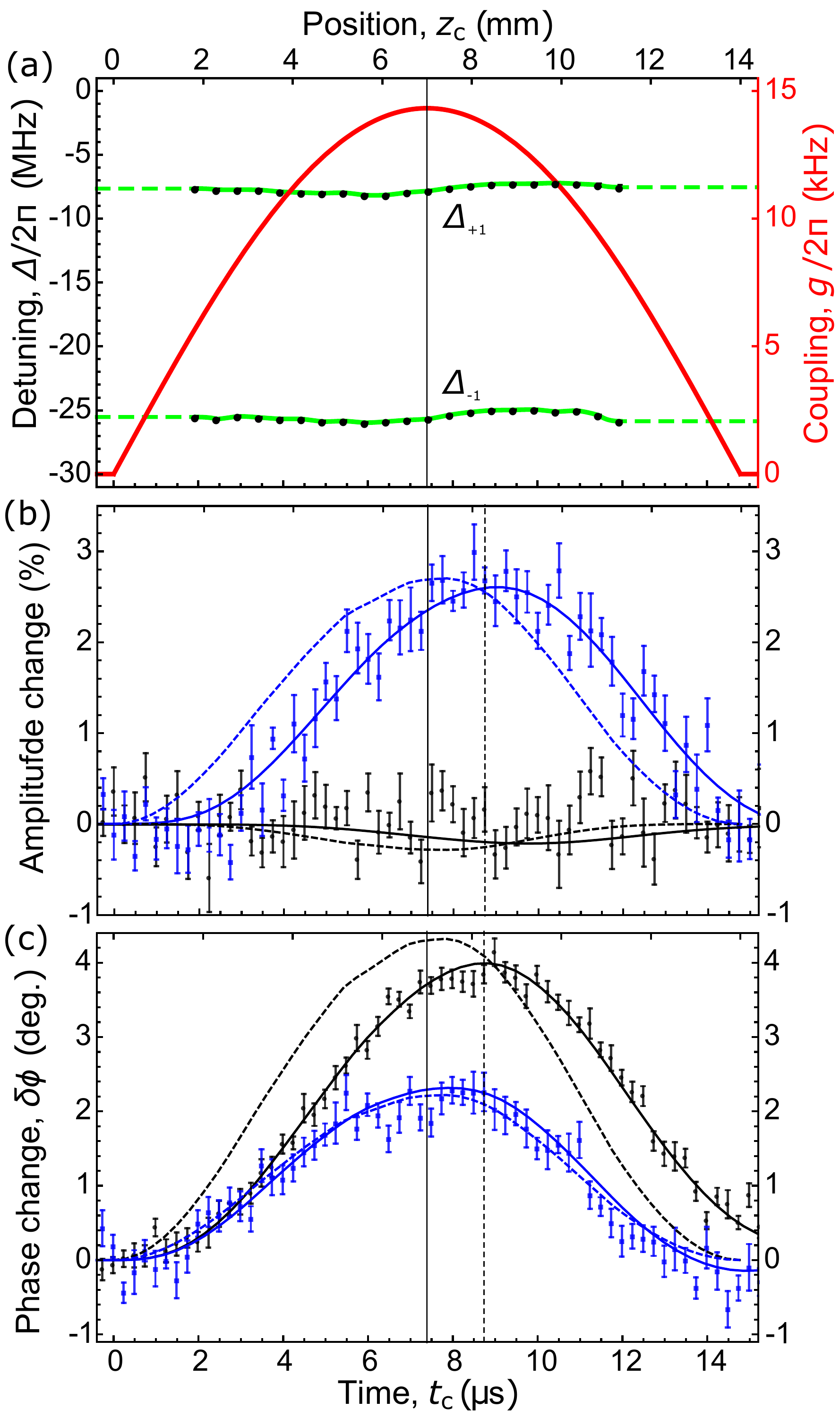} 	
\caption{Dependence of (a) the parameters of the cavity dispersive shift, (b) the amplitude change and (c) the phase change on the position of the atom ensemble in the cavity ($z_{\mathrm{c}}$, top axes) and the time since the ensemble entered the cavity ($t_{\mathrm{c}}$, bottom axes). (a) Atom-cavity detunings $\Delta_{\pm 1}$ (black dots, green fitted curves, left axis) and single atom coupling $g_1$ (red curve, right axis). (b,c) Amplitude and phase changes for a microwave probe frequency on resonance (black) and detuned by $+\kappa/2$ (blue). The data are averaged over $5.5\cdot 10^4$ repetitions and the error bars represent $\pm$ the standard uncertainty. The full lines are fits to the data points using the model (see text), dashed lines represent the response neglecting the cavity photon lifetime. The vertical full and dashed lines indicate the cavity center time $t_{\mathrm{cen}}$ and $t_{\mathrm{max}} = t_{\mathrm{cen}}+2/\kappa$, respectively.}
\label{fig:TimeData} 
\end{figure}	

As the Rydberg-atom cloud propagates through the cavity, the coupling and detuning change as shown in Fig.~\ref{fig:TimeData}(a). We calculated the single-atom coupling from the transition dipole moment ($1431 e a_0$ where $e$ is the elementary charge  and $a_0$ is the Bohr radius) and the microwave electric field distribution and mode volume (calculated using an analytic formula for the field distribution in the rectangular microwave cavity). The single-atom coupling reaches a maximum value of $g_{\mathrm{max}}/2\pi \simeq 14.3 \,$kHz at the cavity center. The detunings are measured by microwave spectroscopy performed within the cavity and are mainly set by the applied magnetic field~\cite{Stammeier2017a} ($7.1\,$G for the data presented in Fig.~\ref{fig:TimeData}). The small variations of the atomic transition frequencies with the position $z_{\mathrm{c}}$ in the cavity are Stark shifts induced by weak, inhomogeneous stray electric fields. The position dependence of the coupling and the detunings, and the propagation of the atom ensemble lead to a time dependence of the dispersive shift $\chi(t)$. 

We calculate the response of the cavity complex transmission ($\mathcal{A} = A e^{i\phi}$ with amplitude $A$ and phase $\phi$) from input-output theory~\cite{Gardiner1985} as:
\begin{equation}
\mathcal{A}(t) = \frac{\kappa}{2} \int_{-\infty}^{t}{\!\!\!\!\!\mathrm{d}t_1 \, \mathrm{exp}\!\left[ \left(i \Delta_{\mathrm{m}}-\frac{\kappa}{2}\right)\left(t-t_1\right) - i\!\int_{t_1}^{t}{\!\!\!\mathrm{d}t_2 \, \chi(t_2)} \right]},
\label{eq:At}
\end{equation}
where $\kappa$ is the energy decay rate of the cavity, and $\Delta_{\mathrm{m}} = \omega_{\mathrm{m}} - \omega_{\mathrm{c}}$ is the detuning between the microwave probe frequency and the cavity. Thus, the transmitted field at time $t$ is averaged over the past evolution of the dispersive shift with weights decaying exponentially on a time scale $\tau_{\mathrm{c}}$.  When the variations of $\chi$ are much slower than the decay time $\tau_{\mathrm{c}} = 2/\kappa$ of the cavity field amplitude, Eq.~\ref{eq:At} simplifies to $\mathcal{A}(t) = 1/[1-2i(\Delta_{\mathrm{m}}-\chi(t))/\kappa]$. This gives the usual dependence of the cavity transmission with $\Delta_{\mathrm{m}}$ as square-root Lorentzian for the amplitude $A$ and arctangent for the phase $\phi$, with an additional frequency shift $\chi(t)$ used in Ref.~\cite{Stammeier2017}. 

We measured the temporal dependence of the dispersive shift of the cavity frequency induced by a cloud of $42$s Rydberg atoms ($P_{\mathrm{s}}=1$, $P_{\mathrm{p},\pm1}=0$) and show, in Fig.~\ref{fig:TimeData}(b,c), the averaged values of the changes in amplitude $\delta A$ and phase $\delta\phi$ with resonant (black) and detuned (blue) microwave probe frequency. On resonance, the transmitted amplitude is only weakly responsive to the dispersive shift because $\delta A$ is measured at the maximum of the Lorentzian line shape and the dispersive shift is small compared to the cavity linewidth. In contrast, the sensitivity of the phase to dispersive shift is maximal on resonance because of the approximately linear variation of the arctangent arround zero. With a detuned microwave probe frequency, the phase change is reduced and the amplitude change becomes significant. The sensitivity of the detection is optimal on resonance where the transmitted power and thus the signal-to-noise ratio are maximum, as indicated by the error bars.

The full lines show the joint fit using Eq.~(\ref{eq:At}) based on the measured decay rate of the cavity photons $\kappa/2\pi = 236(1)\,$kHz, the measured detunings $\Delta_{\pm}$, and the calculated coupling strength $g$. In the fit, we numerically integrate Eq.~(\ref{eq:At}) over $10 \, \tau_{\mathrm{c}}$, only use data points at times when the detunings were measured, and assume that the detunings are constant at earlier times. This approximation has negligible effect on the fit because of the exponentially weighted average and the negligible dispersive shift at times for which the coupling strength is low. From the fit, which takes into account the Rydberg-state lifetime, we determine the average atom number $N = 261(1)$ in the center of the cavity. 
The corresponding calculated instantaneous responses are plotted as dashed lines to emphasize the shift in time and the averaging induced by the finite response time $\tau_{\mathrm{c}} = 2/\kappa \simeq 1.35 \,\mu$s of the cavity field. The maximum phase change on resonance is delayed from the cavity center time $t_{\mathrm{cen}}$ to approximately $t_{\mathrm{max}} = t_{\mathrm{cen}}+\tau_{\mathrm{c}}$ and its value is slightly reduced by the averaging. The detuned response experiences less delay because the field decay rate increases with increasing detuning from resonance. 

We also used the fit of the cavity transmission to determine the time at which the atoms enter the cavity, i.e. the origin of $t_\mathrm{c}$. This preliminary fit, performed before the fit presented above, uses the full time-span of the recorded data and returns the fitted parameter of the origin of $t_\mathrm{c}$. The fitted value has an uncertainty lower than $0.05\,\mu$s. It is consistent with two other independent but less precise measurements, one using the measured atomic velocity and the position of the cavity in the setup, the other based on the measured evolution of the coupling strength from Rabi oscillations induced by a microwave probe tone resonant with an atomic transition.

\FloatBarrier
\section{C. Trueness of the detection}
\label{app:Trueness}

We provide in this section a detailed estimation of the trueness~\cite{BIPM2012} of the measurement of the number of Rydberg atoms. The trueness of the measurement depends on the systematic errors that can arise from the parameters of the dispersive shift, i.e. the calculation of the coupling strength and the measurement of the detunings, and from the approximations of the model: the dispersive approximation, the approximation of a point-like cloud of atoms and the approximation of non-interacting Rydberg atoms.

Regarding the coupling strength, as explained in Section B, its value is calculated from the transition dipole moment and the microwave electric field distribution of the cavity mode. We calculated the transition dipole moment by finding eigenstates of the Rydberg atom hamiltonian and calculating the overlap integral of the dipole between the states of interest~\cite{Zimmerman1979}. This calculation uses the precisely determined values of the quantum defects of helium~\cite{Drake1999} and yields a relative uncertainty on the transition dipole moment typically on the order of $10^{-4}$~\cite{Zimmerman1979}. The microwave electric field distribution of the cavity is approximated by the analytic formula of a rectangular cavity~\cite{Pozar2011}. By performing finite-element simulations of the microwave field distribution of the cavity~\cite{Stammeier2017a}, we conclude that the square of the coupling constant, to which the dispersive shift is proportional, deviates by approximately $-1\%$ from the value of the analytic formula of a perfect rectangular cavity. We thus have a relative systematic error of approximately $+1\%$ on the measured atom number. 

The detunings between the cavity frequency and the atomic frequencies are measured by microwave spectroscopy in the cavity, see Fig.~\ref{fig:TimeData}(a). The mean relative uncertainty of this measurement is $0.7\%$, which can thus induce a relative systematic error on the number of Rydberg atoms of $\pm 0.7\%$.

The approximations of the model can also lead to systematic errors. The first approximation is the dispersive approximation, which is valid when the collective coupling $g\sqrt{N}$ is small compared to the detunings $\Delta_{\pm1}$. Indeed, the dispersive approximation only considers the second order in $g\sqrt{N}/\Delta_{\pm1}$ of a series expansion of the energy of the cavity-like eigenstate. The next terms in this series expansion is the forth order. By calculating this term, we find that the relative error of the dispersive approximation is given by $g^2 N (\Delta_{+1}^2+\Delta_{-1}^2)/(\Delta_{+1}^2\Delta_{-1}^2) \simeq 0.2\%$, where the calculated value uses the maximum value of the coupling $g_{\mathrm{max}}$, $N=600$ and the measured average detunings of Fig.~\ref{fig:TimeData}(a). In general for the atom number detected in our experiment, the dispersive approximation thus leads to an underestimation of the number of atoms with a relative systematic error of $-0.1(0.1)\%$.  

Our model also assumes a point-like atom cloud. In reality, the cloud has a measured longitudinal Gaussian size of $\sigma_z\simeq 0.6\,$mm (for a laser pulse duration of $0.5\,\mu$s) and a transverse one of $\sigma_x\simeq 0.3\,$mm. Because the cloud size is much smaller than the characteristic scale of the variations of the coupling and the detunings (on the order of half the cavity lenght $\sim 7\,$mm), we can reasonably make the point-like approximation to simplify the calculation such that the fit converges within a realistic time span. This approximation leads a to slight underestimation of the atom number, because the spatial distribution of the atoms reduces the collective coupling. We calculated this effect for a cloud with the measured size given above and we found that, in this case, the atom number is underestimated with a relative systematic error of $-3.3\%$. 

Our model additionally assumes that interactions between Rydberg atoms are negligible. The Rydberg atoms can have strong dipole-dipole interactions that are a useful feature for quantum information processing. However, the dipole-dipole interactions decrease rapidly with the interactomic distance R and increase strongly with the effective principal quantum number $n^{*}$ of the Rydberg state. For two atoms in the same state, the Van der Waals interaction energy is $h C_6/R^6$. For the case of the helium triplet $42$s state, the $C_6$-coefficient is $C_6 \sim 36\,$MHz.$\mu$m$^-6$ (calculated from the value for $70$s state given in Ref.~\cite{Zhelyazkova2016} and the $n^{*\, 11}$-dependence with the effective principal quantum number). For two atoms in two different states with an allowed dipole transition, the resonant dipole-dipole interaction is dominated by the exchange term $h C_3/R^3$. For the case of the helium triplet $42$s and $42$p states, the $C_3$-coefficient is $C_3 \sim d^2 / (4 \pi \epsilon_0 h)\simeq 2\,$GHz.$\mu$m$^-3$ as calculated from the value of the transition dipole moment $d \simeq 1431 e a_0 $, the vacuum permittivity $\epsilon_0$ and the Planck constant $h$. In our experiment, we find from the measured cloud dimensions and Rydberg-atom number that the density at the cloud center is about $4\cdot 10^5\,$atoms.cm$^{-3}$. Thus, the average spacing between Rydberg atoms is on the order of $75\,\mu$m in the center of the cloud. At this separation, the interaction between two helium triplet $42$s atoms leads to a frequency shift of about $0.2\,$mHz and the interaction between a $42$s atom and a $42$p atom results in a frequency shift of about $5\,$kHz. Thus, the interactions between Rydberg atoms are negligible compared to the detunings between the cavity and atomic frequencies, which are on the order of $10\,$MHz. 

In conclusion, by adding all the previously described effects, we obtain an approximate relative systematic error of $-2.4(0.8)\%$ characterizing the trueness of our detection of the number of atoms. We note that the systematic errors and their associated uncertainties do not constitute intrinsic limitations of the presented measurement method and could be improved in future implementations.  

By using the small difference observed between the independent measurements of the MCP sensitivity of $2.12(1)\cdot 10^{-2}\,\mathrm{V}\cdot\mathrm{ns}/$atom with the cavity detection and $2.07(4)\cdot 10^{-2}\,\mathrm{V}\cdot\mathrm{ns}/$atom with the MCP calibration, we can also estimate the relative systematic error of the Rydberg atom number detected with the cavity to $-2.3(1.8)\%$, which is consistent with our previous estimate.

\FloatBarrier
\section{D. Measurement of p-state sublevel populations}
\label{app:ProbaP}

\begin{figure}[t] \includegraphics[width=80mm]{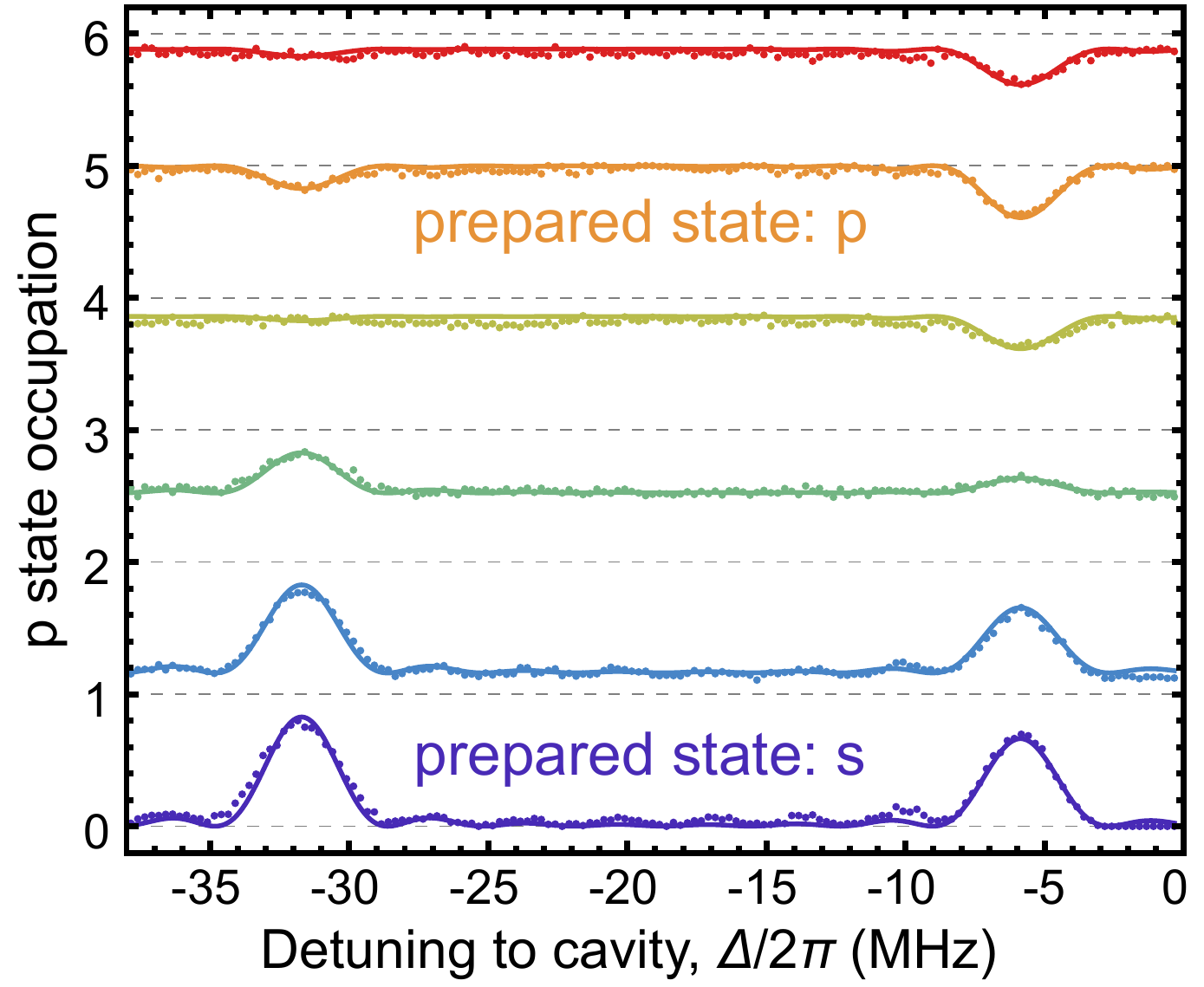} 	
\caption{Intra-cavity spectroscopic measurement of different superpositions between s and p states prepared before the atoms enter the cavity. The occupation of p state in the cavity after the spectroscopy microwave pulse is plotted as function of its frequency. The data points and fits correspond to linearly increasing Rabi frequencies (from 0 to $\Omega/\Omega_{\pi} \simeq 1.24$, shifted vertically by multiples of $1$ for improved visibility) of the preparation microwave pulse which changes the polarization of the ensemble before the atoms enter the cavity. We determine the p-state sublevel occupations from the fitted curves (cf text).}
\label{fig:ProbaP} 
\end{figure}

To measure the intra-cavity detunings presented in Fig.~\ref{fig:TimeData}(a), we measure the population of the p-state induced by a strong microwave pulse applied to an s-state cloud inside the cavity. The pulse is applied through the cavity input port and its power is adjusted to compensate for the Lorentzian dependence of the intra-cavity power on the detuning between the pulse frequency and the cavity frequency. A typical spectrum is presented in Fig.~\ref{fig:ProbaP} (dark blue data set). The resonance peaks at low and high frequency correspond to transitions from the s state to p $m_l = -1$ and p $m_l = +1$ states, respectively.

This spectroscopic measurement can also be realized for a set of superpositions of s and p $m_l = +1$ states, prepared using a microwave pulse applied before the atoms enter the cavity. In Fig.~\ref{fig:ProbaP}, we present the spectra reflecting the occupation of the p state after the microwave spectroscopy pulse, for linearly increasing Rabi frequency $\Omega$ of the drive pulse, shifted by $1$ along the vertical axis for better visibility.
When the atoms are prepared in the p$,\, m_l = +1$ state (orange curve), the baseline of the spectrum corresponds to $P_{\mathrm{p}} = 1$ when the frequency is off resonance. We observe a resonance peak at high frequency, corresponding to transfer from the prepared p$,\, m_l = +1$ state to the s state, as expected, but also a resonance peak at low frequency, corresponding to a transfer from the p$,\, m_l = -1$ state. The presence of the latter indicates that a significant part of the population of the ensemble is in this sublevel of the p state. We think that the depolarization of the ensemble between the preparation and the cavity originates from  stray electric fields at the cavity-entrance hole. Indeed, if stray electric fields induce dc Stark shifts of the atomic levels on the order of the Zeeman shift, but with a different quantization axis, the p-state sublevels are mixed, which results in a depolarization of the ensemble.  As the cavity field polarization does not allow us to drive the transition between the s and p$,\, m_l = 0$ states, we cannot observe directly the population in this sublevel. Nevertheless, by comparing the population transfer observed for p$,\, m_l = \pm 1$ states to the population transfer from the s state spectrum, we determine the fraction $P_{\mathrm{p,\pm 1}}$ of atoms in sublevels $m_l = \pm 1$. We perform a joint fit of the spectra presented in Fig.~\ref{fig:ProbaP} which includes:  

(i) the transfer probability $\sin^2(\delta t \, \Omega /2 )$ to the p,$\, m_l = +1$ state induced by the resonant preparation pulse with Rabi frequency  $\Omega$ and pulse duration $\delta t \simeq 0.4 \, \mu$s; 

(ii) the radiative decay of the s and p states in the time after the preparation pulse and before the spectroscopy pulse; 

(iii) the fraction $P_{\mathrm{p,\pm 1}}$ of atoms in sublevels $m_l = \pm 1$ in the cavity;

(iv) the transfer probability
\begin{equation}
\frac{\Omega_{\mathrm{i},\pm 1}^2}{\Omega_{\mathrm{i},\pm 1}^2 + \Delta_{\mathrm{i},\pm 1}^2} \sin^2(\frac{\delta t_{\mathrm{i}}}{2} \sqrt{\Omega_{\mathrm{i},\pm 1}^2 + \Delta_{\mathrm{i},\pm 1}^2})
\end{equation}
of the transitions induced by the intra-cavity spectroscopy field characterized by the Rabi frequencies $\Omega_{\mathrm{i},\pm 1}$, the detunings $\Delta_{\mathrm{i},\pm 1}$ to the atomic transitions and the pulse duration $\delta t_{\mathrm{i}} \simeq 0.3 \, \mu$s. Here, the fitted Rabi frequencies have a $20\%$ difference resulting from a small deviation of the cavity transmission from a pure Lorentzian at large detuning from resonance, which leads to an imperfect compensation of the intra-cavity power (see difference of peak height of the s-state spectrum).

The fit then yields the p-state-sublevel occupations $P_{+1} = 0.61(3)$ and $P_{-1} =0.20(3)$ for the $m_l = +1$ and $m_l=-1$ states in the cavity. These values are used in Fig.~2(b) of the main text to explain the dispersive shift observed with a preparation of the ensemble in superposition states.

\FloatBarrier
\section{E. Precisions of single-shot detections}
\label{app:Precision}

The precision (characterized by the statistical standard deviation $\sigma$~\cite{BIPM2012}) of the non-destructive detection of the Rydberg-atom number is governed by the precisions of the phase detection and the dispersive shift, which are both dependent on the number of photons populating the cavity. In this appendix, we first discuss how these dependences lead to the choice of the critical photon number as the optimal microwave probe power for the single-shot detection. Second, we provide a comparison of the relative precisions of the cavity-transmission and ionization detections.

The single-shot power signal-to-noise ratio is $R_{\mathrm{S/N}} = n_{\mathrm{c}} \kappa_{\mathrm{out}} \tau_{\mathrm{i}} / n_{\mathrm{noise}} $ which is proportional to the number of photons in the cavity $n_{\mathrm{c}}$, to the coupling of the output port $\kappa_{\mathrm{out}}/2\pi \simeq  150 \,$kHz, to the integration time $\tau_{\mathrm{i}} \simeq 6.2\,\mu$s and inversely proportional to the effective noise photon number $n_{\mathrm{noise}} \simeq 23$ of the detection chain, which is dominated by the noise of the first amplifier (high-electron-mobility transistor amplifier from Cosmic Microwave Technology, model CIT118). The precision of the phase determined by a single measurement on resonance is given by $\sigma_\phi = 1 / \sqrt{R_{\mathrm{S/N}}}$ in the limit $R_{\mathrm{S/N}} \gg 1$ that is valid for the range of parameters discussed here. We measure the change in phase $\delta\phi$ resulting from the dispersive shift. We thus require a reference measurement for the phase without Rydberg atoms in the cavity. We acquire this reference measurement after the atoms left the cavity to avoid long term drifts of the phase~\cite{Stammeier2017}. We perform this measurement with an integration time $\tau_{\mathrm{i,r}} = \alpha \, \tau_{\mathrm{i}}$ which we can choose longer than the integration time $\tau_{\mathrm{i}}$ to improve the single-shot phase precision (empirically, we chose $\alpha = 4$). The precision of the phase change is characterized by 
\begin{equation}
\sigma_{\delta\phi} = \frac{1}{ \sqrt{R_{\mathrm{S/N}}}} \sqrt{1+(\frac{2 \chi}{\kappa})^2 + \frac{1}{\alpha}} \ ,
\end{equation}
where the first two terms in the square root originate from the main phase measurement (with reduced power due to the dispersive shift and the Lorentzian resonance curve) and the last term is caused by the reference measurement. If we additionally consider $\chi \ll \kappa$, as is the case in the experiments which we describe here, we can simplify to $\sigma_{\delta\phi} \simeq \sqrt{\beta / R_{\mathrm{S/N}}}$ using $\beta = 1 + 1/\alpha$.

\begin{figure}[t] \includegraphics[width=80mm]{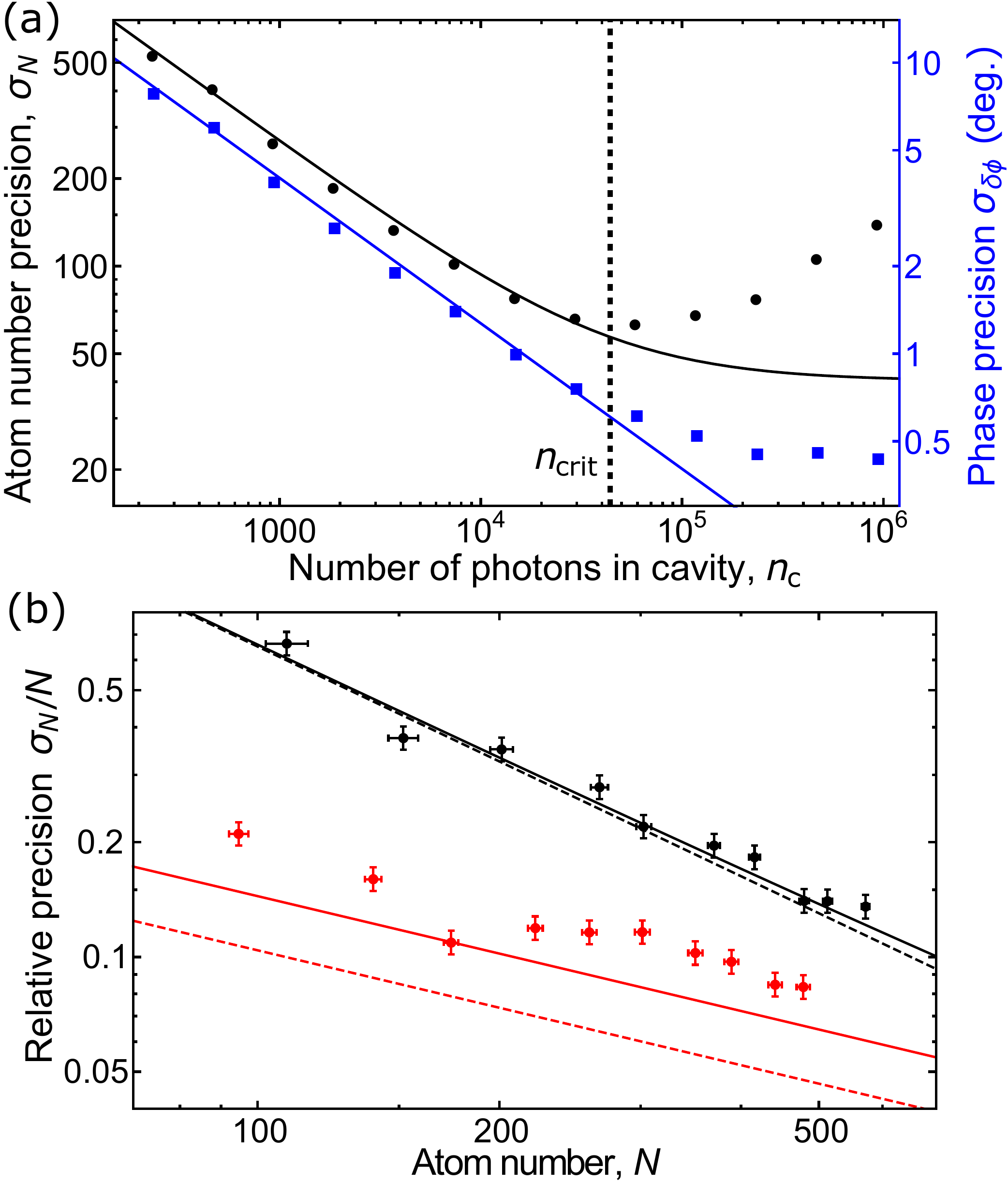} 	
\caption{(a) Atom number precision $\sigma_N$ (black) deduced from the precision of the phase measurement $\sigma_{\delta\phi}$ (blue). The deviation from the calculated dependence (lines) at large photon numbers $n_{\mathrm{c}}$ is due to the finite precision of the phase detection by the FPGA-based device (limited by its intrinsic noise). (b) Comparison of the relative precisions of the Rydberg-atom number detections for the destructive (red) and non-destructive (black) measurements (same data set as Fig.~3(c) of the main text). Atom numbers differ slightly because of the radiative decay of the Rydberg states. Error bars indicate $\pm$ the standard uncertainty. The dashed and full lines show the expected precisions without fluctuations of the prepared atom number and with fluctuations by assuming a Poisson distribution, respectively.  }
\label{fig:Uncertain} 
\end{figure}

The observed phase change is $\delta\phi \simeq - \arctan[2 \chi(n_{\mathrm{c}})/\kappa]$ with $\chi\left(n_{\mathrm{c}}\right)=\chi_0/\sqrt{1+n_{\mathrm{c}}/n_{\mathrm{crit}}}$, if we neglect the small averaging effect of the cavity photon lifetime. This equation can be reversed to obtain the atom number as a function of the phase change and the number of cavity photons. If we additionally consider the case $\chi \ll \kappa$ and also a single coupled transition for simplicity, we obtain for the precision of the atom number by a measurement of phase change:
\begin{equation}
\sigma_{N} =  \frac{\kappa}{g} \sqrt{\frac{\beta \, n_{\mathrm{noise}}}{2\, \kappa_{\mathrm{out}} \, \tau_{\mathrm{i}}}} \sqrt{1 + \frac{n_{\mathrm{crit}}}{n_{\mathrm{c}}}} \ .
\label{eq:sigmaN}
\end{equation}
The last term in this expression shows that increasing the number of photons $n_{\mathrm{c}}$ produces a significant gain in the atom-number precision only up to the critical photon number $n_{\mathrm{crit}}$.

We have measured the precision of our single-shot detection of the phase change (see Fig.~\ref{fig:Uncertain}(a)). The data agree well with the theoretical curve $\sqrt{\beta / R_{\mathrm{S/N}}}$. $\sigma_N$ decreases as $1/\sqrt{n_{\mathrm{c}}}$, up to a point where our precision becomes limited by the intrinsic noise of the FPGA-based device and saturates. On the same graph, we show the corresponding calculated precision of the atom number, which uses the fitted relation between the phase change and the number of photons from Fig.~3(a) of the main text. The theoretical curve thus shows the $\sqrt{1 + n_{\mathrm{crit}}/n_{\mathrm{c}}}$ dependence of the precision. The reduction of the dispersive shift combined with the saturation of the precision of the phase measurement leads to an increasing atom-number uncertainty. This effect only becomes significant above the critical photon number where we reach our best single-shot precision of $\sigma_N \simeq 65$. 

\

When comparing the relative precision of the cavity and ionization detection methods in Fig.~\ref{fig:Uncertain}(b), we first observe that the relative precisions of the two methods are on the same order of magnitude. 

We also see a clear difference in scaling with the atom number $N$ between the two methods. Indeed, besides being non-destructive, the cavity-transmission detection possesses the advantage, over the detection by field ionization, of a more favorable scaling of the relative precision with the atom number $N$.  On one side, the relative precision of the cavity detection scales as $1/N$, as shown by the black dashed curve; because the precision of the atom number depends on the precision of the phase detection, which is independent of the number of atoms (in the limit $\chi \ll \kappa$).
On the other side, for the ionization detection, the MCP is subject to the shot noise caused by its finite detection efficiency ($\eta\simeq 55\%$) and to the fluctuations of the single-event avalanche signal (measured relative standard deviation $\sigma_{\mathrm{a,rel}}\simeq 0.38$). These stochastic processes induce fluctuations of the MCP signal proportional to $\sqrt{N}$. Thus, the relative precision of the ionization detection scales as $1/\sqrt{N}$. Indeed, in the limit where the number of atoms is large ($N \gg 1$), the central limit theorem and the convolution of Gaussian probability density functions allows us to derive an approximate expression for the relative atom-number precision 
\begin{equation}
\sigma_{\mathrm{N,MCP,rel}}\simeq\sqrt{\frac{1}{N}\left(\frac{\sigma_{\mathrm{a,rel}}^2}{\eta} +\frac{1}{\eta} -1 \right)} \ ,
\label{eq:relerrMCP}
\end{equation}
shown as the red dashed curve in Fig.~\ref{fig:Uncertain}(b). 
The measured relative standard deviations for both methods also include the fluctuations of the number of prepared atoms. By assuming that the fluctuations of the number of atoms follows a Poisson distribution and adding this effect to the previous calculations, the theoretical expectations (full lines) match reasonably well the measured precisions. 

Even if the destructive detection by ionization is more precise for the atom numbers produced in our experiment, the cavity detection would outperform it with larger Rydberg ensembles.

\end{document}